\documentclass[sigconf,nonacm,balance=false]{acmart}
\usepackage{popets}

\usepackage{hhline}

\usepackage{subcaption}

\usepackage[shortlabels]{enumitem}

\newcommand{\point}[1]{\par\vspace*{1ex}\noindent\textbf{#1.}~}

\newcommand{\bolditem}[1]{\par\vspace*{1ex}{\noindent\textbf{#1}}}

\newcommand{\lowdetail}{\textit{low-detail}}

\newcommand{\highdetail}{\textit{high-detail}}

\newcommand{\minorsection}[1]{%
  \vspace{1ex}%
  {\bfseries #1}%
  \vspace{0.5ex}%
  \par%
}

\newlist{questions}{enumerate}{2}
\setlist[questions,1]{%
    label=\textbf{RQ\arabic*}.,
    ref=\textbf{RQ\arabic*},
    left=2pt,
    itemindent=0pt,
    topsep=2pt, 
    partopsep=0pt,
    parsep=0pt,
    itemsep=1pt
}
\setlist[questions,2]{%
    label=(\alph*),
    ref=\thequestionsi(\alph*),
    left=2em,
    itemindent=0pt,
    topsep=0pt,
    partopsep=0pt,
    parsep=0pt,
    itemsep=0pt
}

\newtheorem{definition}{Definition}[section]

\begin{document}

\title[A Confidential Computing Transparency Framework for a Comprehensive Trust Chain]{A Confidential Computing Transparency Framework for a Comprehensive Trust Chain}

\author{Ceren Kocaoğullar}
\affiliation{%
  \institution{University of Cambridge}
  \country{}}
\authornote{Corresponding author: ceren.kocaogullar@cl.cam.ac.uk}

\author{Tina Marjanov}
\affiliation{%
  \institution{University of Cambridge}
  \country{}
}

\author{Ivan Petrov}
\affiliation{%
  \institution{Google}
  \country{}
}

\author{Ben Laurie}
\affiliation{%
 \institution{Google}
 \country{}
}

\author{Al Cutter}
\affiliation{%
  \institution{Google}
  \country{}
}

\author{Christoph Kern}
\affiliation{%
  \institution{Google}
  \country{}
}

\author{Alice Hutchings}
\affiliation{%
  \institution{University of Cambridge}
  \country{}
}

\author{Alastair R. Beresford}
\affiliation{%
  \institution{University of Cambridge}
  \country{}
}

\renewcommand{\shortauthors}{Kocaoğullar et al.}

\begin{abstract}
Confidential Computing enhances privacy of data in-use through hardware-based Trusted Execution Environments (TEEs) that use attestation to verify their integrity, authenticity, and certain runtime properties, along with those of the binaries they execute.
However, TEEs require user trust, as attestation alone cannot guarantee the absence of vulnerabilities or backdoors.
Enhanced transparency can mitigate the reliance on naive trust.
Some organisations currently employ various transparency measures, including open-source firmware, publishing technical documentation, or undergoing external audits,
but these require investments with unclear returns.
This may discourage the adoption of transparency, leaving users with limited visibility into system privacy measures. 
Additionally, the lack of standardisation complicates meaningful comparisons between implementations.
To address these challenges, we propose a three-level conceptual framework providing organisations with a practical pathway to incrementally improve Confidential Computing transparency.

To evaluate whether our transparency framework contributes to an increase in end-user trust, we conducted an empirical study with over 800 non-expert participants. 
The results indicate that greater transparency improves user comfort, with participants willing to share various types of personal data across different levels of transparency.
The study also reveals misconceptions about transparency, highlighting the need for clear communication and user education. 
\end{abstract} 


\maketitle

\section{Introduction}\label{sec:intro}

Confidential Computing is a privacy-enhancing technology that aims to keep data secure and private while it is \emph{in-use} by using hardware-based Trusted Execution Environments (TEEs).
This relatively new approach extends the long-standing ability of protecting data \emph{at-rest} using encryption, and \emph{in-transit}, using cryptographic protocols, such as TLS.
Several TEE implementations are available, such as Intel SGX \cite{intelsgx} and TDX \cite{inteltdx}, AMD SEV-SNP \cite{amdsev}, and Arm CCA \cite{armcca}.
Commercial cloud platforms, including Google Cloud \cite{googlecloudcc}, Microsoft Azure \cite{azurecc}, and Amazon Web Services \cite{awsnitro} have made Confidential Computing widely available.

Using TEEs provides Confidential Computing with distinctive properties compared to other privacy-preserving computation techniques like homomorphic encryption and secure multiparty computation. 
Specifically, TEEs provide confidentiality for both data and code in execution through techniques such as hardware isolation and memory encryption.
TEEs also protect the integrity of the data and code throughout execution.
This makes Confidential Computing a versatile tool that can be used independently or complement other privacy methods, such as homomorphic encryption. 
Therefore, Confidential Computing is an essential component in building a foundation for comprehensive private computation.

A fundamental challenge in Confidential Computing is that it requires a certain level of user trust.
\textit{Attestation}, a key feature of Confidential Computing \cite{confidential2022terminology}, provides information about the authenticity, integrity, and certain runtime properties of a TEE through validating evidence signed by the hardware.
The evidence contains relevant measurements of the Trusted Computing Base (TCB) \cite{confidential2022terminology}, which is the collection of firmware and software that the security of a system depends on \cite{latham1986orangebook}.
However, while attestation verifies that the intended software is running on an intended TEE, it does not guarantee the absence of vulnerabilities or backdoors within a Confidential Computing system. 
Therefore, unless the user builds and installs the entire software stack that runs on top of the TEE by themselves, and the hardware is correct and reviewable, attestation alone does not provide enough evidence to the user to trust a Confidential Computing system (\S\ref{sec:transparency-requirement}).

Transparency offers a complementary approach to overcome these inherent limitations of attestation.
While transparency does not ensure security, as evidenced by notable vulnerabilities found in open-source software \cite{xzbackdoor, heartbleed, log4shell, poodle, shellshock}, it offers a significant improvement over the alternative of naive trust. 
Transparency allows community scrutiny, which can lead to quicker detection and resolution of issues, thereby promoting greater accountability. 
Transparency is not limited to source code:
Certificate Transparency is now the standard method for monitoring TLS certification \cite{laurie2013rfccertificate, laurie2014certificate}, inspiring similar initiatives such as Key Transparency \cite{melara2015coniks, googlekeytransparency, whatsappkeytransparencyblog, whatsappkeytransparencywhitepaper} and Binary Transparency \cite{al2018contour, lins2023mobile}.

Currently, companies are adopting varied transparency ap\-proach\-es to reduce users' reliance on naive trust in Confidential Computing.
Critical software components are increasingly open source, including AMD SEV firmware \cite{amdfirmware}, Intel TDX modules \cite{inteltdxmodule}, and ARM Trusted Firmware \cite{armtrustedfirmware}.
Apple PCC \cite{applepcc2} also made code for certain key security and privacy components (such as attestation, verification and logging) open source, and provided a Virtual Research Environment where reviewers can boot and inspect software.
The AWS Nitro architecture underwent a recent audit, and the findings have been made public \cite{awsaudit1, awsaudit2}.

However, while open-source code and professional audits contribute to reducing the need for user trust, the quality and comprehensiveness of these transparency measures vary significantly.
For example, Intel TDX modules are open source with extensive documentation and reproducible builds \cite{inteltdxmodule}, whereas the AMD SEV firmware lacks even basic documentation \cite{amdfirmware}.
Similarly, AWS Nitro audits do not provide any technical guarantee that the audited component designs match those deployed in production environments \cite{awsaudit1}.
The resulting lack of standardisation makes it difficult to meaningfully evaluate and compare transparency provided by different implementations.
Moreover, implementing transparency can be costly with no immediate clear benefit to organisations. 
As a result, while larger organisations may make some efforts to provide transparency, smaller ones may be discouraged by the substantial challenges involved.

To address this challenge, we propose a Confidential Computing Transparency framework, which systematises the landscape of possible solutions (\S\ref{sec:transparency-levels}).
Our framework defines an end-to-end trust chain using methods such as open-source code, reproducible builds and provenance or endorsement statements.
These methods are underpinned by digital certificates and verifiable transparency logs to allow user attestation. 
Our framework has three transparency levels, each defined by different agents of transparency, who can be first-party, third-party or community reviewers.
The tiered structure provides organisations with a practical pathway for incrementally enhancing the transparency of complex Confidential Computing systems in real-world deployments. 
Similar to maturity models like C2M2 \cite{c2m2}, CMMC \cite{cmmc}, and SLSA \cite{linuxfoundation0}, this tiered structure allows organisations to progressively improve their transparency practices as their capabilities and resources allow.
Although focused on Confidential Computing, the framework’s principles may also extend to broader hardware and binary transparency contexts.

To understand the impact of the transparency levels of our framework on user trust, we conducted a comprehensive user study with over 800 non-expert participants.
The study focused on two key aspects: (1) how different levels of transparency affect end-user trust in Confidential Computing, and (2) what types of data end-users are willing to share with Confidential Computing systems at different transparency levels.
Our results indicate that the trust end-users place in Confidential Computing systems correlates positively with the increasing levels of our framework, with the type of data used playing a role as well.

Additionally, we ran two treatments of the study---one where we give a more abstract, less detailed explanation of Confidential Computing, transparency and the proposed framework; and one where we provide more detail.
We find that a more detailed explanation and a discussion of common misconceptions leads to higher comfort with increasing levels of transparency and reduces the rate of misconceptions.

Our framework describes the implementation required to adopt different levels of transparency, and shows that transparency can be implemented gradually with a tangible increase in user trust at each step.
\textbf{Our main contributions are summarised as follows:}
\begin{itemize}
    \item We establish key concepts related to transparency in Confidential Computing, including its definition, significance, scope, beneficiaries and facilitators (\S\ref{sec:defining-transparency}).
    \item We propose a three-level Confidential Computing Transparency framework, bridging technical security in TEEs and user trust through reviewer accountability and a robust trust chain (\S\ref{sec:transparency-levels}).
    Our framework offers increasing transparency levels, adaptable for complex needs and limitations of real-world systems.
    \item We test the impact of our framework on user trust with a large-scale user study with over 800 participants (\S\ref{sec:user-study}).
    We show that increased transparency positively impacts trust, and that users are comfortable sharing different types of sensitive data at different transparency levels. 
    \item  We identify a number of common misconceptions related to transparency. 
    We find that adding an additional one minute of explanation in our introductory video reduces the rate of these misconceptions by at least half.
    Moreover, users are more comfortable with increased transparency when their understanding of Confidential Computing and why we need transparency increases (\S\ref{sec:user-study}).
\end{itemize}
\section{Defining transparency}\label{sec:defining-transparency}

In this section, we answer five important questions about transparency in Confidential Computing: what is it and why is it important (\S\ref{sec:transparency-requirement}), what is subject to transparency (\S\ref{sec:scope}), who stands to benefit from it (\S\ref{sec:beneficiaries}), and who facilitates it (\S\ref{sec:agents}). 
We defer answering the question of how transparency can be achieved to \S\ref{sec:transparency-levels}.

\subsection{Necessity of transparency}\label{sec:transparency-requirement}
Attestation is a tool that provides information on the trustworthiness of a TEE.
It often involves a hardware-signed proof of some information about the origin and current state of a TEE \cite{confidential2022technical}.
There are three types of claims that may be derived from this signed proof:

\begin{enumerate}
    \item \textit{\textbf{Authenticity:}} 
    Attestation allows users to verify the origin of the TEE and the workload running inside it.
    It involves verifying that the TEE has been produced by the expected manufacturer, and that it runs the expected workload.
    \item \textit{\textbf{Integrity:}} 
    Attestation can provide users with claims about whether software components like firmware, bootloader, operating system and workload, have been manipulated.
    \item \textit{\textbf{Runtime configuration:}} 
    Attestation can additionally provide partial insights into the runtime configuration of the current state of the TEE, such as the execution mode.
\end{enumerate}

Although attestation is essential for building trust in the authenticity and integrity of a TEE \cite{confidential2022technical}, it has limitations in providing comprehensive security assurance. 
Notably, it cannot detect vulnerabilities or backdoors within attested components. 
For instance, a backdoor in the TEE’s firmware or hardware Trusted Computing Base (TCB) could allow unauthorised access to user data without affecting the measurable authenticity or integrity verified by attestation, and the limited runtime information included in attestation evidence.

A transparent approach with thorough review processes is crucial for uncovering vulnerabilities or backdoors. 
Such reviews can include inspecting the source code of critical binaries, evaluating the tools used in generating the binaries from the source code, and conducting comprehensive assessments of the system's architecture (\S\ref{sec:scope}). 
The review methodologies may involve manual inspections, automated testing, formal reasoning (see \S\ref{sec:automated-certifiers}), and other systematic techniques.
Embracing this kind of transparency can mitigate hidden threats, empower users and stakeholders to make informed decisions, and ultimately provide a high level of assurance about the security of a Confidential Computing system. 

\subsection{Scope of transparency}\label{sec:scope}
Transparency in Confidential Computing must extend beyond the TEE. 
Communication channels with a TEE, where users load private data and code, must be secured, for example, by using a cryptographic protocol. 
Similarly, when multiple TEEs (e.g. those on a CPU and a GPU) collaborate, the communication between them must also be secure. 
Therefore, we consider the entire end-to-end Confidential Computing systems rather than focusing only on TEEs.

In an end-to-end Confidential Computing system, we classify components that could jeopardise the confidentiality or integrity of user data if compromised as \textit{sensitive components}. 
These include the TEE TCB for security and integrity (e.g. the components that process plaintext user data, generate or store keys), as well as endpoints and protocols used in TEE-to-user or TEE-to-TEE communication.
While transparency should ideally extend to all sensitive components, its benefits depend on three key prerequisites:

\begin{enumerate}
    \item\textit{\textbf{Reviewability:}} 
    Security insights about a component should be achievable through inspection.
    This goal is inherently subjective and must be considered case-by-case.
    For example, the ideal approach to achieve reviewability for a software component is granting reviewers access to the source code.
    However, even with source code access, nuances exist \cite{thompson1984reflections}; access to the documentation, architecture, reference manuals, or expert guidance might impact reviewability.
    The same applies to granting access to the commit history versus a snapshot of source code. 
    In cases where providing source code access is not feasible, other forms of reviewability can also be achieved, for example by executing a binary within a confined environment and analysing its behaviour, as Apple did with PCC \cite{applepcc2}.
    While the highest level of reviewability is ideal, practical constraints may call for a best-effort approach.
    \item\textit{\textbf{Certifiability:}} 
    It should be possible to generate and publish a digitally signed statement for a reviewed component.
    Certifiability is crucial for maintaining the integrity and transparency of the review process, potentially enhancing the quality of reviews, aiding in compliance with legal and regulatory requirements, providing evidence in disputes, etc.
    Forms of certifiability vary, such as providing a signed statement for reviewed source code or, in the case of a frequently fine-tuned machine learning model, certifying the architecture instead of the weights.
    \item\textit{\textbf{Attestability:}}
    While transparency addresses limitations of attestation in providing security assurance, attestability itself serves as a prerequisite for transparency.
    This paradoxical need arises because without attestability, users cannot confirm whether a specific instance corresponds to a reviewed version of a component, and certifiers cannot be unequivocally held accountable for their assessments (\S\ref{sec:transparency-requirement}).
    Therefore, transparency cannot enhance a component's security assurance in an externally verifiable way without attestability.
\end{enumerate}

We call the sensitive components that meet all three criteria as \textit{transparency-enhanced} sensitive components.

\medskip

\point{Hardware transparency}
A notable category of sensitive components that may not fit the \textit{transparency-enhanced} description is hardware. 
While hardware designs can be open source, like OpenTitan \cite{opentitan}, reviewed, and even formally verified \cite{kern1999formal}, there appears no scalable method for verifying that any piece of hardware matches a particular design, with the same technical guarantees as software. 

Methods like supply chain auditing and monitoring can provide some level of insight into hardware integrity.
Apple PCC counters targeted hardware attacks by using high-resolution imaging in the manufacturing chain, and auditing hardware at the data centres under third-party oversight \cite{applepcc}.
Another approach can be inspecting random hardware samples to verify that they correctly correspond to the design.
However, these methods fall short of the certainty provided by reproducible software builds.
As a result, \textit{reviewability} and \textit{certifiability} of hardware components present significant challenges.
The same applies to \textit{attestability}, as current attestation protocols focus on software elements like firmware and drivers, offering little or no assurance about hardware.

These fundamentally restrict a user's ability to assess whether the hardware can support Confidential Computing; let alone its correctness.
Given these challenges, this paper focuses on the software components of Confidential Computing systems.
If these challenges are resolved, our transparency framework can be used for hardware as well.
For now, the security properties gained by our transparency framework (\S\ref{sec:transparency-levels}) rest on the correctness of the hardware.

\point{Non-sensitive components}
Transparency for \textit{non-sensitive components}, which do not threaten the confidentiality or integrity of user data if compromised, requires establishing a well-defined trust boundary. 
This trust boundary acts as a sensitive component, protecting integrity and confidentiality of data.
Transparency can then be used to externally validate that this trust boundary prevents the non-sensitive components from compromising confidentiality and integrity of data. 
Assurance in this context can be achieved through methods like hardware isolation or memory encryption.
If the trust boundary is transparency-enhanced, non-sensitive components can be excluded from the transparency discussions.
Otherwise, non-sensitive components should be treated as sensitive and included in the transparency efforts.

\subsection{Agents of transparency}\label{sec:agents}
A Confidential Computing system achieves transparency when users can review its sensitive components directly or delegate this responsibility to trusted parties.
These reviewers must remain accountable to users by providing verifiable evaluations and justifying their decisions.
Accountability begins with establishing a verifiable link between a review and its reviewer.
We implement this link through digitally signed review certificates, which uniquely identify both the reviewed component and the reviewer.
Therefore, we call the reviewers \textit{certifiers}, who act as agents of transparency by providing traceable evaluations.

This work only focuses on establishing the link between reviews and reviewers, with broader mechanisms for detecting and enforcing accountability left for future exploration. 
While institutional reviewers benefit from legal frameworks and governance structures, achieving accountability for individual reviewers requires structured systems, including verifiable digital identities, reputation mechanisms, and transparent review histories.

\textbf{\textsc{Traits:}} Each certifier has traits relating to their \textit{methodology} and \textit{motivation}. 

In terms of \textit{methodology}, certifiers can be either reporting or alerting.
A certifier's methodology influences the certificate content.

\bolditem{Reporting certifiers} may uncover both the strengths and shortcomings of the component under review.
The code owner may provide a signed public review plan to guide certifiers, similar to Protection Profiles in Common Criteria \cite{commoncriteria}.
They issue certificates, that are similar to Security Assessment Reports (SAR), including review scope and findings.
Format of these certificates can be standardised by the code owner or an independent entity.
Users may choose their own trusted reviewers, similar to the open-source Rust code review system \texttt{cargo-vet} \cite{rustcargovet}, or form a web of trust, as in \texttt{cargo-crev} \cite{rustcargocrev}, another Rust review system.
Also similar to \texttt{cargo-vet} \cite{rustcargovet}, users can define policies to ensure that their trusted reviewers assess the code according to specific criteria. 

\bolditem{Alerting certifiers} focus on finding bugs and vulnerabilities.
They issue certificates detailing the discovered issues, including information such as the vulnerability type, root cause, impact description, and public references---similar to CVE records \cite{cveprogram}.
These certificates are useful for urging the code owner to fix the issues and holding them accountable if they fail to do so.
After a fix alerting certifiers must assess and issue a follow-up certificate that includes their opinion on the solution. 
The number and content of alerting certificates, certifier credibility, and code owner's response serve as trust signals.

\begin{figure}
\includegraphics[width=0.48\textwidth]{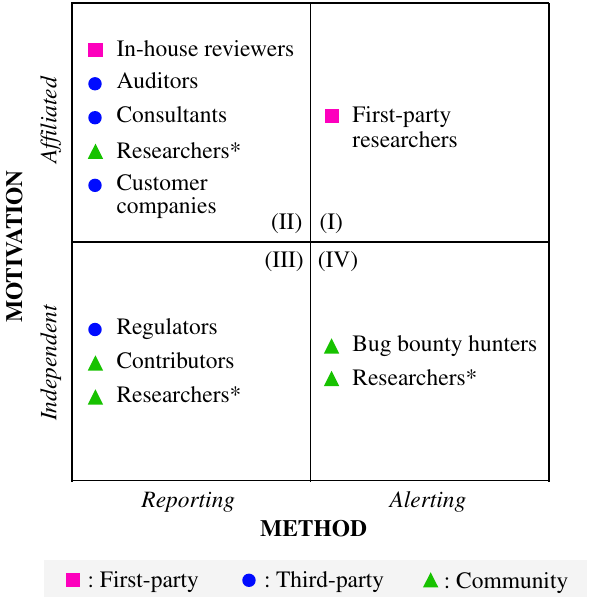}
\caption{
Graph showing all certifier trait combinations, with each quadrant including examples from the three reviewer categories where applicable. 
An asterisk (*) indicates that researchers may have varying motivations and methods, such as (II) those funded by code owners, (III) volunteer researchers in community efforts, and (IV) those focused on finding vulnerabilities and publishing papers.
}
\label{fig:reviewer-quadrant}
\end{figure}

In terms of \textit{motivation}, certifiers can either be independent, or they may be affiliated to the code owner.
\bolditem{Independent certifiers} have no vested interests in the code owner's outcomes and are not responsible to them.
They may have a greater commitment to ensuring the component's quality and security, but can also be driven by self-interest: independent researchers seeking publication, open-source contributors who wants to gain experience, individuals affiliated with competing organisations who aim to highlight weaknesses in the code owner’s product, or bug bounty hunters motivated by financial rewards.

\bolditem{Affiliated certifiers} have formal or informal associations with the code owner through contractual agreements, employment, partnerships, financial ties, etc. 
While affiliation introduces potential impact on their assessments, they are incentivised to maintain high standards due to accountability for their certificates and the risk to their reputation.
A structured reputation system tracking review quality, consistency, and peer feedback can further encourage thorough and truthful assessments.

\smallskip

\textbf{\textsc{Categories:}}
We categorise certifiers into three groups based on their access methods to the reviewed components: \textit{first-party}, \textit{third-party}, and \textit{community certifiers}.
As Figure \ref{fig:reviewer-quadrant} shows, the categories of reviewers are orthogonal to their traits.

\bolditem{First-party certifiers} are internal to the \textit{code owner}, the entity responsible for producing the reviewed component. 
They are typically involved in the component's development or oversight and have the authority to address issues directly. 
All first-party certifiers are affiliated, and most likely reporting in methodology.
However, there are also examples of alerting first-party certifiers, including first-party security research teams like Google Project Zero \cite{googleprojectzero}, Microsoft Security Response Center (MSRC) \cite{microsoftsecurityresponsecenter}, and IBM X-Force \cite{ibmxforce}, which identify and disclose vulnerabilities in their own closed-source \cite{CVE-2020-16875, CVE-2021-1223} or open-source \cite{CVE-2020-16009, CVE-2021-29976} software.

\bolditem{Third-party certifiers} are granted exclusive access to the components for review. 
The ideal third-party certifiers are independent experts, including regulators, and to some degree, customer companies. 
For instance, a cloud platform purchasing hardware may review the firmware and drivers, or a company outsourcing software development may review the components to ensure quality.
Customer companies lack full independence due to business dynamics, such as financial ties or concerns about the performance impact on purchased components.
Third-party certifiers may also be affiliated, such as auditors or consultants paid by the code owner.
Third-party certifiers might be obligated to maintain confidentiality of the reviewed components.

\bolditem{Community certifiers} access components through open sourcing and include individuals or groups with relevant expertise, such as academics, independent researchers, bug bounty hunters, and external adopters.
Notable existing community certification programmes include Rust \texttt{cargo-crev} \cite{rustcargocrev} and \texttt{cargo-vet} \cite{rustcargovet}.
Professional research teams such as Google Project Zero \cite{googleprojectzero}, MSRC \cite{microsoftsecurityresponsecenter}, and IBM X-Force \cite{ibmxforce} also act as alerting community certifiers when reviewing code produced by other code owners \cite{msrcthirdparty1, msrcthirdparty2, projectzerothirdparty1, projectzerothirdparty2}.

\subsection{Beneficiaries of transparency}\label{sec:beneficiaries}

We identify four types of users who may benefit from transparency:

    \bolditem{End users} have their data processed within a Confidential Computing system, whether on a remote resource or their own devices. 
    Their primary benefit is maintaining both privacy and integrity of their data.
    \bolditem{Application developers} execute code within Confidential Computing systems, where their priority may be protecting the confidentiality of their code, their customers' data, or both.
    This user category includes application developers using cloud-based services such as Google Cloud Confidential Computing \cite{googlecloudcc}, Microsoft Azure Confidential Computing \cite{azurecc}, and AWS Nitro \cite{awsnitro}.
    It also includes users who use end-device hardware to run code, such as operating systems, hypervisors, or other low-level components. 
    \bolditem{Application providers} execute third-party code within Confidential Computing systems.
    Their primary concern is ensuring the security and trustworthiness of the code they run.
    \bolditem{Platform providers} include cloud providers, software-as-a-service providers, CPU manufacturers as platform providers for executing code in enclaves, Android with Private Compute Core \cite{marchiori2022androidpcc} or Android Virtualisation Framework (AVF) \cite{avf}, and other entities offering infrastructure or services for Confidential Computing. 
    Their main objective is to establish and maintain a trustworthy Confidential Computing system, ensuring the integrity and security of the systems and services provided to end-users and developers.

\begin{figure}
\includegraphics[width=0.47\textwidth]{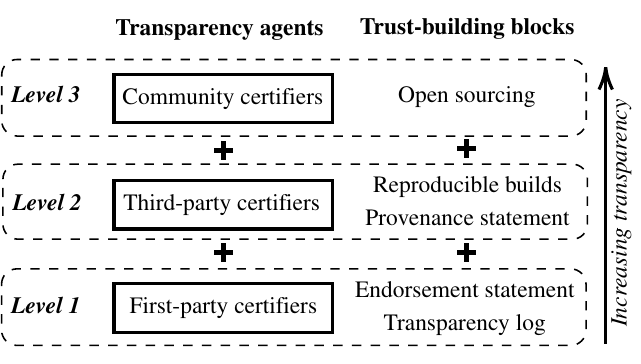}
\caption{
A depiction of the three transparency levels in our framework, showing the agents and trust-building blocks needed to achieve them. 
Each level builds on the previous one by adding more trust-building elements.
}
\label{fig:transparency-levels}
\end{figure}

\section{Confidential Computing Transparency}\label{sec:transparency-levels}

Following the reasoning and discussions we provide in \S\ref{sec:defining-transparency}, we define Confidential Computing Transparency as follows.

\begin{definition}[Transparency]
Transparency in Confidential Computing is the practice of allowing certifiers (as categorised in \S\ref{sec:agents}) to examine the security properties of the \textit{sensitive}, \textit{reviewable}, \textit{certifiable}, and \textit{attestable} (as defined in \S\ref{sec:scope}) components of a Confidential Computing system.
These certifiers should be able to certify their comprehensive assessments in a manner that is publicly verifiable.
\end{definition}

In this section, we present a framework to answer the question of \textit{how} this transparency can be achieved in Confidential Computing.
Our framework acknowledges the diverse transparency needs, limitations, and risk profiles of various system components. 
For instance, some transparency-enhanced sensitive components might not allow open sourcing or third-party certification due to IP restrictions or proprietary information.
This complexity is more pronounced in real-world systems involving multiple stakeholders and components from different vendors.
For example, even a single TEE can include microcontroller units from different vendors, each with dedicated firmware and drivers.

To address these complex requirements and challenges, our framework adopts a layered approach, defining three transparency levels, each incorporating a certifier category described in \S\ref{sec:agents} and a set of trust-building blocks.
As shown in Figure \ref{fig:transparency-levels}, each level builds on the previous one by adding more trust-building blocks and integrating an additional certifier category.
Based on needs and limitations, system designers can apply different transparency levels to different components, though we strongly advise aiming for the highest transparency level achievable.

\subsection{Level 1 (L1)}\label{sec:transparency-level-1}

\point{Requirements} 
The foundational level of transparency is achieved through engaging \textit{first-party certifiers} as accountable reviewers.
This level relies on the following trust-building blocks to ensure transparency:
\begin{description}[leftmargin=1em,labelindent=1em]
    \item[\textit{Endorsement statement}] is a digitally signed statement by affiliated and reporting first-party certifiers, including a unique identifier of a binary (e.g. hash), as defined and used by Google's open-source project Oak \cite{oak}.
  It can also include the time of issuance and the validity period.
  By issuing an endorsement statement for a specific binary, the certifier confirms that the binary was generated from specific source code and endorses its use in production for the validity period. 
  After the validity period, the certifier may issue a new statement or passively revoke the existing one by taking no action (see \S\ref{sec:revocation}).
    \item[\textit{Verifiable transparency log}] is a publicly available append-only ledger maintained on a trusted infrastructure by an independent party, separate from the code owner or the certifiers \cite{laurie2014certificate, melara2015coniks, chuat2015efficient, hu2021merkle}. 
    Transparency logs do not inherently guarantee truthful and accurate operations, so additional measures are required to ensure their integrity and consistency, as discussed in \S\ref{sec:custodians}.
\end{description}

\point{Process}
At this transparency level, the review and certification must happen prior to the binary's release.
There is one exception to this rule: if first-party researchers identify a vulnerability in the binary after its release, they may issue alerting certificates.
The first transparency level is achieved through the following steps.
\begin{enumerate}
    \item First-party certifiers review the source code associated with the sensitive components, and approve it for release.
    \item The code owner builds a binary from the source code. 
    To ensure the integrity of the build process, it is essential that both the code owner and the first-party certifiers have established trust in the build toolchain and the infrastructure used. 
    This trust is ideally built through direct review of the tools and processes. 
    If direct review is not feasible, it may be reasonable to implicitly trust open-source tools, or tools widely used across the company and maintained by a dedicated team.
    The code owner can then release the binary, e.g. by distributing it to cloud servers, end devices, or publishing it to an app store.
    \item \label{step:level-1-endorsement-statement}
    First-party certifiers generate an endorsement statement that they sign with their private signing keys (separately or using a multisignature scheme \cite{itakura1983public}).
    They publish this endorsement statement on a transparency log.
    \item \label{step_level-1-alerting}
    If first-party researchers discover a vulnerability after the binary's release, they generate a signed alerting certificate as described in \S\ref{sec:agents}, and publish it on the transparency log.
    \item \label{step:level-1-log-check}
    After receiving a binary, the user must verify that the transparency log contains a valid endorsement statement. 
    This involves: (I) confirming the endorsement's binary hash\footnote{If the binary is running on a server where the user is the client (as opposed to the user's end-device) the user must obtain the attestation evidence of the binary to retrieve the correct hash value.} and validity period, and (II) verifying the endorsement's signature(s) using the certifiers' public keys, unless relying on third-party monitors (\S\ref{sec:custodians}).
    In Binary Transparency, this verification is typically done by an \textit{auditor}, usually represented as client-side software \cite{al2018contour}. 
    Inclusion proofs can be used for efficiently verifying that an endorsement statement is added to the transparency log (\S\ref{sec:inclusion-proofs}). 
    The user (or auditor) must also periodically check the log for any alerting certificates, which may prompt them to stop using the binary or take relevant measures.
\end{enumerate}

\point{Discussion}
At this transparency level (L1), beneficiaries cannot directly see the attested components, but generating and publishing endorsement statements holds the code owner accountable for their released binaries. 
This ensures the code owner cannot disown a particular release.
Furthermore, this transparency level ensures that any insider or outsider attacks on the released binary do not go unnoticed, as new transparency log entries or their absence are detectable.
This achieves `Signature Transparency', a concept implemented by Sigstore's Rekor \cite{rekor} and Sigsum \cite{sigsum}.

L1 also guards against covert targeted attacks, where an attacker serves a specific user a targeted malicious binary by allowing users to verify they are served the same binary as everyone else. 
However, this system is not immune to more overt, coordinated attacks. 
Even if the transparency log operations are monitored for correctness as described in \S\ref{sec:custodians}, an adversary can manipulate the timing of population updates. 
For instance, the adversary might delay visibility of transparency log updates for certain IPs or intentionally cause out-of-band checks for update availability to fail for specific IPs. 
This way, the adversary can potentially create split-view release streams with `good' and `bad' binary versions. 
Even so, this transparency level can alert vigilant observers to anomalous activities like unusual release patterns or duplicate version numbers, serving as a deterrent and early warning mechanism against targeted attacks.

As all first-party certifiers are affiliated (see \S\ref{sec:agents}), reviewers in this transparency level are limited to the first and second quadrants of the certifier diagram (Figure \ref{fig:reviewer-quadrant}).
In other words, there is no variety of motivation among certifiers at this transparency level.

\subsection{Level 2 (L2)}\label{sec:transparency-level-2}

\point{Requirements}
The second transparency level is achieved by introducing \textit{third-party certifiers} as accountable certifiers.
First-party certifiers also participate at this level, performing reviews and issuing certificates similarly to how they operate in L1. 
This ensures that code owners remain accountable for the binaries they release. 
In addition to the trust-building blocks used in the first transparency level, the second level introduces the following ones. 
\begin{description}[leftmargin=1em,labelindent=1em]
    \item[\textit{Reproducible builds}] refers to a collection of tools and techniques that are used to ensure that every build for the same source code consistently generates the same bit-exact binary output \cite{reproduciblebuilds}. 
    \item[\textit{Provenance statement}] includes all the necessary configuration details for building the source code into a specific binary, such as what toolchain, commands and flags to use \cite{oak}.
    It is provided by the code owner to certifiers alongside the component under review.
\end{description}

\point{Process}
The build, release, and first-party review process of the binary happens as described in L1 (\S\ref{sec:transparency-level-1}). 
To attain L2 transparency, the following additional steps must be followed.
At this transparency level, the review and certification can happen before or after the binary's release.

\begin{enumerate}
    \item Third-party certifiers review the source code of the sensitive components.
    The source code must be reproducibly buildable, enabling each certifier to build it and generate the same binary as other reviewers (see \S\ref{sec:trusted-builder} for the alternative). 
    By comparing the hash of this binary with the signed certificates in the transparency log, certifiers can ensure they have reviewed the same source code as others.
    \item \label{step:level-2-self-build}
    Each certifier uses the provenance statement to learn the build configurations for the component and independently compiles the source code into a binary on trusted infrastructure. 
    Certifiers must establish trust in the build toolchain instructed by the provenance statement, achievable by applying transparency principles to the toolchain itself (\S\ref{sec:binary-transparency}).
    \item \label{step:level-2-endorsement-statement}
    Each third-party certifier generates a signed certificate (either reporting or alerting as described in \S\ref{sec:agents}) for the self-built binary, and publishes it on a transparency log.
    \item \label{step:level-2-log-check}
    Once the binary is received on an end-device (or the attestation evidence if the binary runs on a server), the user should check if a valid endorsement statement for it exists on a transparency log, as in L1 Step \ref{step:level-1-log-check}.
    Unlike endorsement statements, the existence and validity of certificates from third-party certifiers is not enough to provide trust to the beneficiaries; their contents matter (\S\ref{sec:agents}).
\end{enumerate}

\point{Discussion}
The second transparency level (L2) significantly improves trust by incorporating both affiliated and independent third-party reviewers, unlocking the third quadrant and expanding the diversity of reviewer motivations (Figure \ref{fig:reviewer-quadrant}).
From a practical standpoint, incorporating third-party certifiers may introduce potential delays in the review process, particularly in dynamic production environments where code updates are frequent. 
This creates a trust dilemma, especially in scenarios where a vulnerability is detected. 
The code owner may need to decide whether to quickly implement a patch without waiting for certification, or follow the full review process and delay until certification is received.

To mitigate these time-related challenges, one potential solution might be implementing a transparent source control mechanism to allow certifiers to review only the code changes.
An alternative approach is issuing certificates after the binary is released.
Although this might mean users are not immediately assured that third-party certifiers have vetted the binary processing their data, this mechanism disincentivises the code owner from releasing malicious or subpar code.
To implement this approach, the code owner should include an explicit promise in the endorsement statement at Step \ref{step:level-2-endorsement-statement} that there will be a third-party audit certificate to follow. 
This promise may also specify a timeline for certification, e.g. \texttt{to\_be\_certified\_by: <date>}.
Having a signed explicit promise like this eliminates ambiguity and allows for more automated verification of the promise at Step \ref{step:level-2-log-check}.

Making the source code reproducibly buildable can be challenging, or the reviewers might find it impractical to build the binary themselves. 
An alternative approach is using a trusted builder (see \S\ref{sec:trusted-builder}).
However, this option comes with certain caveats, including the need for the certifiers and beneficiaries of transparency to trust the builder. 
    
\subsection{Level 3 (L3)}\label{sec:transparency-level-3}

\point{Requirements}
The third and highest level of transparency employs \textit{community certifiers} as accountable reviewers. 
As in L2, community certifiers do not replace the first-party and third-party certifiers from the previous levels.
L3 must include first-party certifiers, and ideally, it should also incorporate third-party certifiers.
In addition to the trust-building blocks used by the other levels, this transparency level introduces a new one:
\begin{description}[leftmargin=1em,labelindent=1em]
    \item[\textit{Open sourcing}] in this context refers to making the source code of a component publicly accessible on a platform, which the user does not need to trust.
\end{description}

\point{Process}
In this level, each community certifier follows the same process described in L2 (\S\ref{sec:transparency-level-2}).

\point{Discussion}
This level of transparency significantly improves openness and accountability, driven by principles of public accessibility and community involvement. 
As we discuss in \S\ref{sec:agents}, signed certificates provide the foundation for accountability by creating verifiable links between reviews and their reviewers.
While certificates establish identity, complete accountability for community certifiers requires broader mechanisms like reputation systems, peer audits, and enforcement protocols.

Similar to L2 (\S\ref{sec:transparency-level-2}), this transparency level may not always guarantee real-time verification. 
Open sourcing, while a powerful tool for transparency, does not guarantee that all security issues will be detected immediately, as evidenced by past vulnerabilities in large-scale open-source projects with extensive communities including Heartbleed \cite{heartbleed}, POODLE \cite{poodle}, Log4Shell \cite{log4shell}, and Shellshock \cite{shellshock}.
Nonetheless, open sourcing remains a crucial step towards minimising the need for user trust.
Moreover, unlike traditional open sourcing, this transparency level requires reviewers to certify their assessments, serving as an additional trust signal (\S\ref{sec:agents}).

Availability of source code substantially improves the ability of independent alerting certifiers to analyze code for defects (\S\ref{sec:agents}), fully unlocking the last and fourth quadrant of the certifier diagram (Figure \ref{fig:reviewer-quadrant}).
To incentivise such certifiers, the code owner may set up bug bounty programmes.
The code owner may also set up coordinated vulnerability disclosure mechanisms to allow the alerting reviewers to report issues to the code owner and giving it some time before issuing a public certificate.
In this case, if the issue is fixed before it is disclosed to the public, the certifiers may still issue their certificate after a patch has been released, similar to how first-party security research teams file CVE records after patch releases.

\subsection{Revocation}\label{sec:revocation}
Revocation is an essential part of the transparency framework, as the majority of software is ultimately revoked. 
Two main reasons to revoke an endorsement statement or a binary are:

\begin{description}[leftmargin=1em,labelindent=1em]
    \item[\textit{Vulnerability identified:}] The code owner or a certifier finds a vulnerability in the endorsed binary. 
    \item[\textit{Log inconsistencies:}]  A monitor, witness (\S\ref{sec:custodians}), or auditor (\S\ref{sec:transparency-level-1}) detects anomalies in the transparency log or endorsement statements.
\end{description}

In both of these cases, there are decisions to make about \textit{who} will make the revocation decision and \textit{how} will the revocation be carried out.
For vulnerabilities, the code owner can revoke the endorsement statement.
One way to achieve this is through passive revocation by issuing short-lived endorsement statements and simply not issuing a new endorsement statement for the affected binary. 
This notifies users of the issue, allowing them to stop using the binary if feasible or take appropriate mitigation steps within their organisational processes.
Alternatively, code owner can actively revoke a statement by publishing its unique identifier on a publicly accessible \textit{certificate revocation list (CRL)} \cite{cooper2008internet, yee2013updates}.

For both revocation reasons above, users can independently choose to stop using a binary based on information they receive from auditors, monitors, or witnesses. 
One way to implement this is through a policy in the client-side auditor software \cite{al2018contour}, similar to the policies in Rust \texttt{cargo-vet} \cite{rustcargovet}.
This policy can alert the client about inconsistencies on a transparency log or the existence of alerting certificates about a binary.
Alternatively, monitors or witnesses can collectively issue global revocation statements, which they can then publish on CRLs.
\section{Additional considerations}\label{sec:variations}
In this section, we describe optimisations and other modifications that can be applied to our transparency framework to accommodate different needs and constraints.

\subsection{Monitoring transparency logs}\label{sec:custodians}
Publicly available transparency logs alone do not ensure truthful and accurate operations. 
Monitors can watch logs for correct behaviour, such as ensuring the log is append-only, hashes are valid, and the log does not present split views to different observers \cite{laurie2021rfccertificate}.
By \textit{gossiping}
\cite{gossipinginct, chuat2015efficient, meiklejohn2020think} their log views, monitors can prevent propagation of inaccuracies, with collective signatures from witnesses mitigating potential attacks on gossiping \cite{syta2016keeping}.

Even if the transparency log behaves correctly, an endorsement statement's presence in a log does not confirm its validity; it only means that the information is discoverable.
Without validating the endorsement statement, the trust relies on the expectation that someone will eventually detect any inaccuracies and raise an alarm, a task we have assigned to clients so far (see L1 Step \ref{step:level-1-log-check} and L2/L3 Step \ref{step:level-2-log-check}).
However, this approach may be impractical due to resource constraints, technical expertise gaps, or large volume of data in transparency logs. 
Clients would also need certifier public keys, which can be challenging to manage.
Instead, beneficiaries of transparency can delegate validation to monitors, which can verify the accuracy of log entries and certificate signatures \cite{laurie2021rfccertificate, al2018contour, syta2016keeping}.
An open-source monitoring mechanism enables individuals to scrutinise its source code and deploy them on their trusted machines, and can leverage the proposed transparency framework.

\subsection{Inclusion proofs}\label{sec:inclusion-proofs}
Users must verify the presence of a valid endorsement statement in the transparency log when they receive a binary or, an attestation evidence for a binary (see L1 Step \ref{step:level-1-log-check} and L2/L3 Step \ref{step:level-2-log-check}).
This can be efficiently done using \textit{inclusion proofs}, similar to those in Certificate Transparency, which verify the inclusion of a certificate on a transparency log \cite{trillian, stark2021certificate, al2018contour}.
To implement inclusion proofs, the transparency log is constructed as a Merkle tree \cite{merkle1987digital} with a tree head signed by the log operator \cite{laurie2021rfccertificate}.
An inclusion proof provides a minimal set of hashes to compute the tree head from a leaf node.
With an inclusion proof and the hash of an endorsement statement (as the leaf node), if one can compute the has of the tree head and also verify its signature, this confirms the endorsement statement's inclusion in the transparency log (given that the client has the correct public key to verify the signature).
To keep inclusion proofs scalable, the log operator periodically issues a signed commitment to the state of the log, or \textit{checkpoint}, which can be used instead of the original tree head \cite{trilliantransparentlogging}. 
This approach reduces the verification complexity from linear to logarithmic in the number of certificates.

While Certificate Transparency also uses inclusion promises, signed commitments guaranteeing future log entries, to address merge delays, these rely on full trust in the transparency log and require additional monitoring \cite{trilliantransparentlogging}. 
Since Confidential Computing transparency can tolerate some publish-time latency, we do not recommend inclusion promises. 
Alternative transparency log designs like Sunlight \cite{sunlight} can help reduce merge delays.

\subsection{Trusted builders}\label{sec:trusted-builder}
In L2 and L3 (\S\ref{sec:transparency-level-2} and \S\ref{sec:transparency-level-3}), the certifiers are responsible for building the binary to ensure the binary they certify matches the reviewed source code.
This task can also be delegated to a trusted builder, a specialised tool for building binaries.
Using a trusted builder presents an important tradeoff: the certifiers are relieved from the responsibility of building the binary themselves, but they need to trust the builder.
Despite this tradeoff, a trusted builder can be particularly useful L3, where community reviewers may lack resources.
Using a trusted builder also eliminates the need for reproducibly buildable source code, simplifying a challenging task for code owners.
However, ideally, source code should still support reproducible builds for independent verification.

When a trusted builder is used, it must generate a \textit{signed} provenance statement for each build and post it on a transparency log. 
This statement confirms the binary was built on a secure platform and provides certifiers with verifiable insight into the build environment. 
This is especially important since the certifiers might not be able to build the binary themselves to compare it with the binary that the trusted builder produced.

To establish trust in a builder, users must first trust the build toolchain. 
This can be achieved by using an open-source build toolchain and applying the transparency principles to it (also see \S\ref{sec:binary-transparency}).
Second, the toolchain must also operate on trusted infrastructure, ideally controlled by an actor unlikely to collude with the code owner.
For instance, Google's open-source Confidential Computing project Oak \cite{oak} has a trusted builder that uses the open-source SLSA build stack \cite{linuxfoundation0} and runs on GitHub, which is currently owned by Microsoft. 
The level of assurance can be further increased by involving multiple independent builders. 
For example, Oak runs a trusted builder instance on Google Cloud in addition to GitHub. 
Another way of increasing the trustworthiness of a builder is to run the build toolchain’s TCB inside a TEE \cite{linuxfoundation2}, which can also be different from the one under scrutiny.
Using a hardened build platform with strong isolation and protection for the signing key, as in SLSA V1.0 \cite{linuxfoundation1}, can also increase trust on the builder.

If a trusted builder is used in  L2 or L3, the release process for the binary is revised as follows:
\begin{enumerate}
    \item The code owner initiates a build of the source code using a trusted builder. 
    \item \label{step:trusted-builder-provenance} The trusted builder generates a signed provenance statement about the binary and the build process, publishes it on a transparency log, builds the binary and returns it to the code owner. 
    If there are multiple trusted builders, then all builders do the same.
    \item The code owner verifies the provenance statement’s signature, and confirms the build process details. 
    For multiple trusted builders, these checks are repeated for each build, and if the code is reproducibly buildable, the owner ensures all binaries are identical.
    \item  If verification succeeds, the code owner generates an endorsement statement referencing the provenance statement. 
    For non-reproducible builds with multiple builders, separate endorsements are issued for each unique binary.
\end{enumerate}

Following this build and release, in L2 or L3, the certifiers review the source code. 
If the source code is not reproducibly buildable, Step \ref{step:level-2-self-build} of L2 and L3 is replaced by the certifier retrieving the binary’s endorsement and provenance statements from the transparency log, validate their signatures, and confirm the binary hash in the endorsement statement matches.
For reproducible builds, certifiers can also independently compile the source code on trusted infrastructure as outlined in Step \ref{step:level-2-self-build} of L2 and L3.

\subsection{Automated certifiers}\label{sec:automated-certifiers}
A certifier can be replaced by automated processes, introducing a fourth certifier category, which we call \textit{automated certifiers}.
An automated certifier can be constructed by building a transparent mechanism that generates formal proofs of security properties of the component.
The transparency of the generation mechanism gives assurance to the beneficiaries (\S\ref{sec:scope}) that the proofs are generated correctly.
This alternative certifier type can be used for achieving transparency in a similar way in L2 and L3, also using a trusted builder as we describe in \S\ref{sec:trusted-builder}.
The level of transparency provided depends on the assurance of the generated proof.
This alternative certifier type can both be seen as an optimisation, since automated proofs may work more efficiently than human reviewers, and as a privacy enhancement, as the code owner can potentially generate proofs that do not reveal proprietary information, e.g. by using zero-knowledge proofs \cite{luo2022proving}.
\section{User study}\label{sec:user-study}

We conducted a user study to examine perceptions of transparency within the Confidential Computing framework. 
Our primary metric is user comfort in a fictitious app. 
Our study varies two factors within participants: the transparency level of each app, and the data type that the app requires. 
The data type was varied because users often view certain types of data as more sensitive than others \cite{chua2021effects, milne2017information, park2018value, skatova2023unpacking}. 

We first conducted the study providing a concise explanation about Confidential Computing and transparency (\lowdetail).
The results of this study indicated some common misconceptions among the participants (\S\ref{sec:user-study-misconceptions}).
Motivated by this, we conducted a second variant (\highdetail) where we increased the level of detail in our explanation, as well as the thoroughness of the comprehension questions.

\subsection{Research questions and method}\label{sec:user-study-method}

We aim to answer two main research questions: 

\begin{questions}
    \item\label{rq-1} How do different transparency levels of the Confidential Computing Transparency framework shape an end user sense of trust?
    \item\label{rq-2} What types of personal data are end users comfortable sharing at different transparency levels?
\end{questions}

In the study, we first introduced the participants to Confidential Computing through a script presented in video format and in text (Appendix \ref{apx:user-study-script}). 
In the first (\lowdetail) run of the study, the script presented information about Confidential Computing and transparency in a relatively abstract manner \footnote{Videos are temporarily available for review, will be made more broadly available when paper is published: \url{ https://tinyurl.com/low-detail}}. 
We took this approach as we wanted to mimic the real-life scenario where non-experts are exposed to technical concepts through high-level summaries, for example in an advertisement. 
In the \lowdetail\ variant, Confidential Computing was presented as a system that protects user data from unauthorised access, including by app developers, illustrated with a vault analogy. 
Participant understanding was assessed with three true or false questions. 

In the \highdetail\ variant of the study, we explain Confidential Computing in more detail, address its shortcomings and provide more information on the review process \footnote{\url{ https://tinyurl.com/high-detail}}. 
Additionally, we highlight that the reviewers do not get access to users' data during review, and that (for the purpose of the study) all reviewers should be considered experts with the same level of access to the system. 
The participants in this variant were presented with ten multiple choice comprehension questions.

In both \highdetail\ and \lowdetail\ variants, after we provided participants with the explanation video and script, we presented them six imaginary virtual assistant apps running on a Confidential Computing system.
Each app required access to a different type of personal data, which we categorised into: (i) social-economic, (ii) lifestyle-behaviour, (iii) tracking, (iv) financial, (v) authenticating, and (vi) medical-health data, drawing from the work of Chua et al. \cite{chua2021effects}.
We selected virtual assistant apps, as they represent a single app type that may require access to any of the six chosen data types.
Each app was randomly assigned a specific transparency level L1, L2, or L3, or no transparency at all.
We described each transparency level using neutral language and labelled all certifier types as \textit{experts}.\footnote{While our framework incorporates first-party certifiers across all transparency levels, we omitted this detail from the study to maintain clear distinctions between levels, avoiding potential confusion and to not imply any hierarchy.}
For each app, we also included an animated GIF illustrating the corresponding transparency level, taken from the informative video.
Participants sequentially saw the apps and rated their comfort levels using the app, and their confidence in the ability of the app to protect their data.

In the final phase of the study, participants rated their comfort level using a virtual assistant app reviewed under different transparency options (L1, L2, L3 and no transparency), regardless of data type. 
They also evaluated their level of concern if their data were leaked or tampered with across various data types, without reference to transparency. 
The goal of this question was to understand the extent to which concerns about privacy are tied to perceptions of risk associated with specific data types.

We also collected information on participants' openness to adopting new technologies and their current habits of using virtual assistant apps to determine how general distrust or familiarity might influence privacy concerns.
We provide a discussion of ethical considerations, user study artefacts, the full study script, and a breakdown of demographics in the Appendix \ref{apx:appendix}.

We conducted the study as a survey using Qualtrics and recruited participants via the Prolific academic platform \cite{palan2018prolific}, as recommended by relevant research \cite{tang2022replication}. 
We obtained ethics approval from (\textit{redacted for anonymity}) review board.
We recruited 817 participants, split roughly equally between the two variants, in small batches between August and November 2024.\footnote{The sample size was calculated through a power analysis based on pilot data. The batches were spread over time and time zones to minimise the effect of time of the day, and day of the week.} 
The median time taken to finish the survey was \(\sim \)15 minutes, and the participants were compensated at a fixed rate of £3 or \(\sim \)£12/h. 
Participants who did not complete the study or were identified as bots were excluded and automatically replaced by Prolific.
Our final analysis after the removals was performed on 758 participants.

\subsection{Quantitative results}\label{sec:user-study-quantitative}

We have two key metrics: participant comfort levels in the core phase of the study, aggregated across all data types, and comfort with the virtual assistant app under given transparency options, irrespective of data type, which we gather from the last phase of the study. 
We report here results based on the first metric for three reasons: (i) results based on both metrics are similar, (ii) the first metric allows us to address both RQs using the same primary outcome, and (iii) tests based on the first metric are more conservative, meaning any results reported hold, and are even stronger with the second metric.\footnote{More specifically, broader variation in comfort levels was observed when participants were exposed to different data types. This variation is largely driven by considerable differences in comfort across various data and transparency combinations. 
Conversely, when data types were not specified, the variability in comfort levels decreased providing a slightly clearer and more homogeneous picture. 
Thus, obtaining significant results based on the first metric was harder, which increases our confidence in the results.}

\begin{figure}[ht!]
    \centering
    \begin{subfigure}{0.99\linewidth}
        \includegraphics[width=\linewidth]{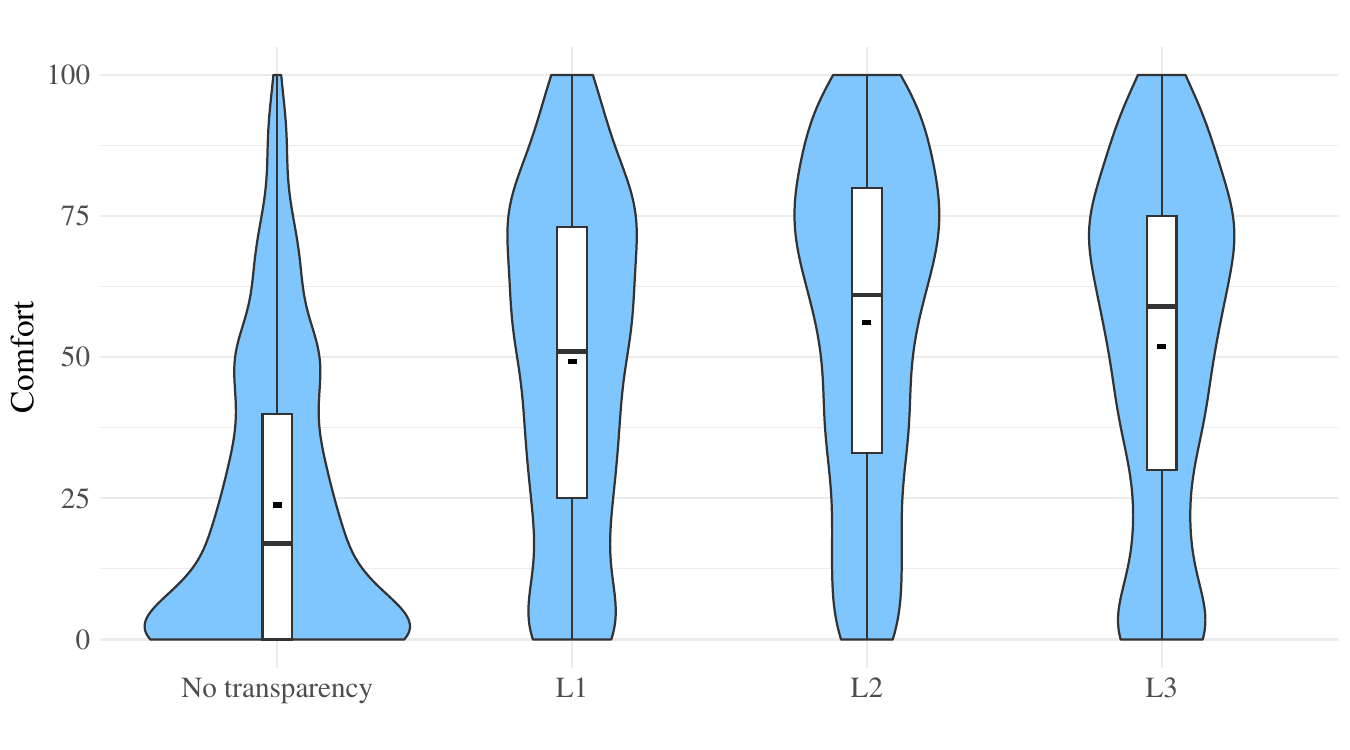}
        \caption{\lowdetail\ variant}
        \label{fig:RQ1-transparency-v1}
    \end{subfigure}
    \begin{subfigure}{0.99\linewidth}
            \includegraphics[width=\linewidth]{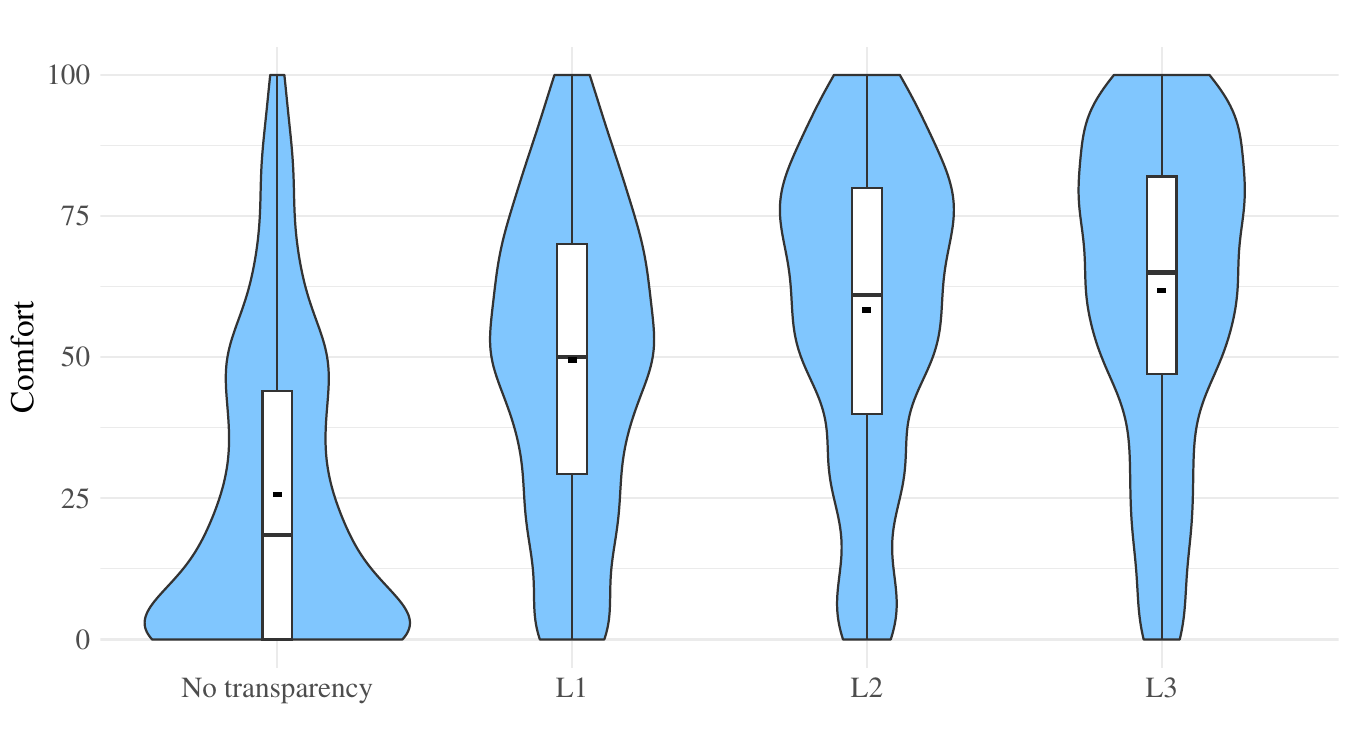}
            \caption{\highdetail\ variant}
            \label{fig:RQ1-transparency-v2}
         \end{subfigure}
  \begin{minipage}{7cm}
    \footnotesize
    The violin plots show the median of each transparency group as a horizontal line segment within each box plot, and the mean as a dot.
  \end{minipage}
    \caption{Violin plots of participants' comfort levels, aggregated over all data types.}
    \label{fig:RQ1-transparency}
\end{figure}

Figure \ref{fig:RQ1-transparency} shows the user comfort levels when sharing their data in the \lowdetail\ and the \highdetail\ variants.
We formally test the effect of transparency on participants' comfort through a two-way ANOVA, and we reject the null hypothesis that all transparency group means are equal (\lowdetail\ variant: $F = 159.992, p < 0.001, N = 2,214$; \highdetail\ variant: $F = 218.023, p < 0.001, N = 2,334$). 

We further investigate which transparency group means differ from each other using Tukey pairwise comparison test. 
Participants in both \lowdetail\ and \highdetail\ variants reported the lowest comfort levels in the absence of transparency. 
In the \lowdetail\ variant, the highest comfort is measured for L2, with L1 and L3 being perceived as roughly equally attractive on average. 
However, in the \highdetail\ variant, we observe a clear ranking of the transparency groups from L1 to L3. 
While the first three transparency options remain relatively stable across the two detail variants, the clear ranking in the \highdetail\ variant is driven by an increase in comfort with L3.

On the individual level, we report the frequency of preferring L3 over L2.
In the \lowdetail\ variant, L3 was preferred roughly equally frequently compared to L2. 
On the contrary, in the \highdetail\ variant L3 is preferred by three quarters of participants.

We additionally examine the effect of data type on participants' comfort regardless of the transparency level.
Participants express highest concern about (and lowest comfort with sharing) their financial and authentication data, followed by tracking and medical data. 
They are least concerned about, and most willing to share, socioeconomic and lifestyle-behaviour data. 
We formally confirm the effect of data types with ANOVA (\lowdetail\ variant: $F = 23.046, p < 0.001, N = 2,214$; \highdetail\ variant: $F = 18.567, p < 0.001, N = 2,334$).
Our findings on participants' baseline concern about different data types largely align with Chua et al. \cite{chua2021effects}.

\begin{figure}[ht!]
\centering
\includegraphics[width=0.48\textwidth]{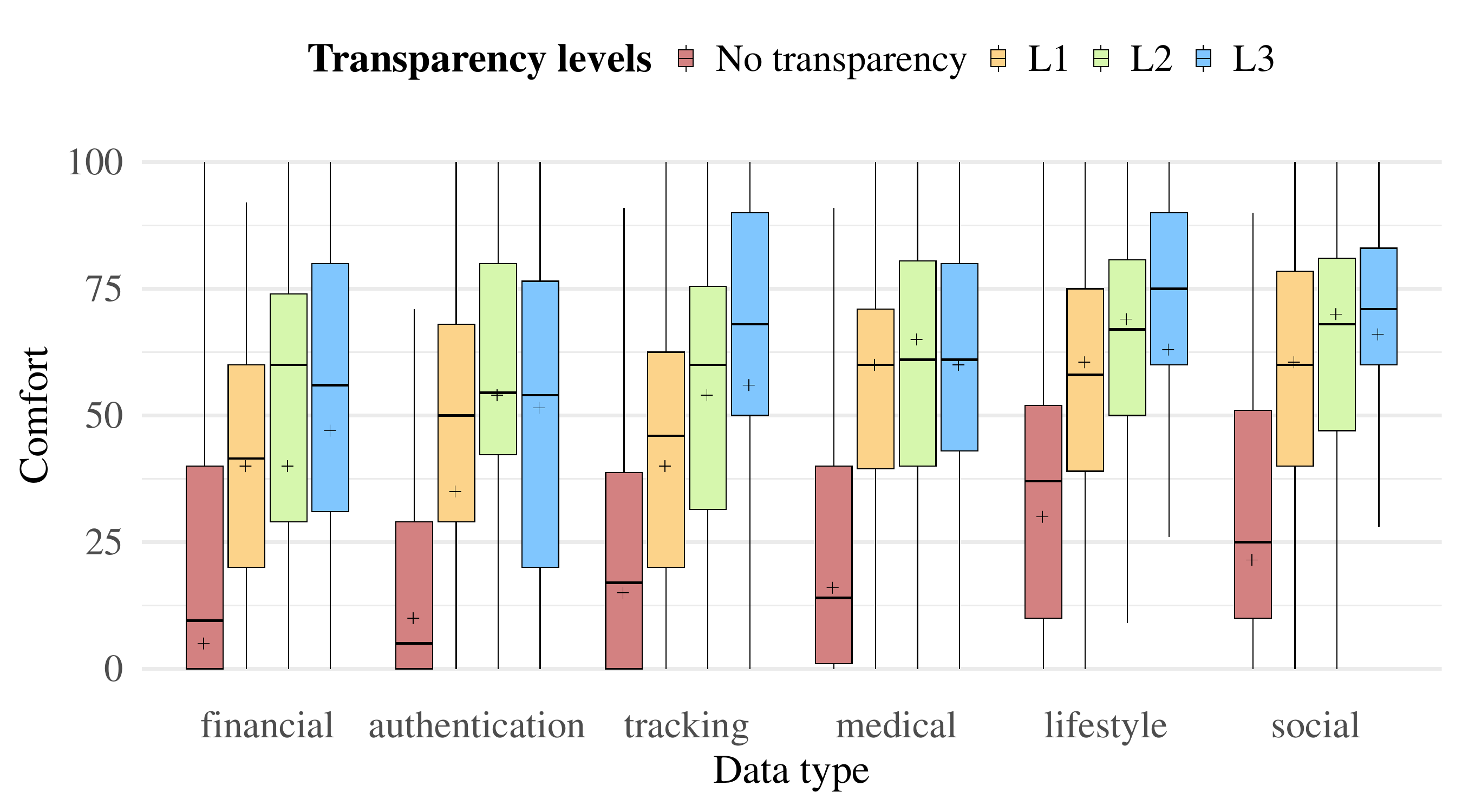}
  \begin{minipage}{7cm}
    \footnotesize
    The boxplots show median comfort levels in the \highdetail\ variant across data types and transparency options. Black dots represent medians in \lowdetail\ variant.
  \end{minipage}
\caption{Boxplots of participants' comfort levels.}
\label{fig:RQ2-transparency_vs_data}
\end{figure}

Upon establishing the primary effect of transparency, we investigate interaction effects between transparency and data type to answer \ref{rq-2} (see Figure \ref{fig:RQ2-transparency_vs_data}). 
We find the previous conclusions from both detail variants generally hold within each data type, with L2 preferred for majority of data categories in the \lowdetail\ variant, and L3 preferred for majority of data categories in the \highdetail\ variant. 
A noteworthy exception is financial data, where the preference between L2 and L3 is reversed in both variants.

Next, we observe the effect of providing more detailed information. 
For less sensitive data types (on the right of Figure \ref{fig:RQ2-transparency_vs_data}), additional information does not change user comfort much or at all, possibly due to ceiling effects as their comfort was quite high in \lowdetail\ variant already. For sensitive data, we see an increase in comfort levels across all transparency levels when more detail is provided.
We also examine data types individually. 
We find that more information does not affect users' comfort of sharing medical data and authentication data. 
In both of these data groups, we see comparable comforts with L1, L2 and L3. 
Focusing again on the \highdetail\ variant, we observe the exact opposite for tracking and lifestyle data, where each higher transparency option is perceived more positively than the previous one. 

Another key insight is the interaction between (i) the type of data processed, and (ii) detail level of the introductory explanation. 
For certain data types (e.g. social, medical and authentication) we find that third-party and community certifiers are roughly equivalent and providing more information about the review process does not matter much. 
On the contrary, for other types of data (e.g. tracking and lifestyle) open sourcing is preferred when users are informed on the benefits of higher transparency.

Finally, we report on exploratory analysis on how comfort levels vary with demographic variables. 
Overall we find that demographic traits do not have a statistically significant effect. 
We report below some noteworthy patterns.
Three participant categories showed greater comfort in sharing data with our fictitious apps: younger participants, participants more willing to explore new technology, or participants from the Middle East and Africa. 
However, comfort levels were not correlated with participant gender.
We also observe some differences in the most preferred transparency option. 
University graduates and participants in arts/humanities, economics and STEM fields prefer L3, whereas participants in social or health sciences and participants with no university education either prefer L2 or are indifferent between the two.

\subsection{User perceptions}\label{sec:user-study-qualitative}
In addition to quantitative data, we aimed to gain a deeper understanding of participants' perceptions and beliefs about transparency levels. 
To achieve this, we included an open-ended question asking participants to elaborate on their rationale behind their choices, allowing them to discuss any or all of the transparency levels freely.
We manually analysed the answers to the open-ended question by first removing non-informative answers that do not address the question (e.g.``I was honest'', ``Personal experience'') or do not provide reasoning relating to any specific transparency level (e.g. ``All seem very safe'', ``I would rather trust my data to machines than to people'').

We started with (738) answers and, after removing (452) empty or non-informative answers, we manually annotated the remaining (286) answers to extract the main themes. 
We used an iterative inductive approach, allowing themes to emerge from the data until saturation is achieved and no new themes are identified. 
Finally, two annotators independently classified the participants' answers into the theme classes. 
The Cohen's $\kappa$ agreement score was 0.82 (almost perfect agreement), with any disagreements later discussed and resolved. 

Participants in both \lowdetail\ and \highdetail\ variants highlighted a similar set of reasons, with roughly equal frequencies. We therefore present them jointly, unless specified otherwise. Participant responses supported the quantitative results, expressing a strong dislike for no transparency (35\% of all annotated answers), believing that unreviewed systems can have vulnerabilities or that the app developer might hide something. 
No answers defended unreviewed systems, however several responses noted reviewing is imperfect as humans make mistakes or that they are satisfied with no review. 

While the consensus is clear for no transparency, the same cannot be said for L1, L2 and L3. 
The main theme emerging from the data is that of objectivity and impartiality (60\% of all annotated answers). 
Specifically, participants believe first-party certifiers cannot be objective and trusted due to conflict of interest, possible pressure from their employer, self-interest and bias. 
External certifiers, on the other hand, are considered more neutral and trustworthy because they are perceived as being less influenced by pressure from the app developer.
Many participants express such views for both community and third-party certifiers (43\%). 
However, a smaller portion believes that only community certifiers can be truly independent. 
This belief is particularly strong among the participants in \highdetail\ variant (42\%), compared to \lowdetail\ variant (14\%).

Participants expressed a range of concerns, sometimes conflicting, in addition to their focus on objectivity. 
Some participants believe first-party certifiers have a better understanding of the app, giving them an advantage in the reviewing process (5\%). 
On the other hand, others feel that third-party certifiers are more precise, and are true experts in the field specialised in reviewing (7\%). 
In the \lowdetail\ variant particularly, some participants \textit{exclusively} addressed third-party certifiers as experts, or perceived them to have better and more exclusive access to the sensitive components, especially when compared to community certifiers (11\% of annotated answers from \lowdetail\ variant).
In contrast, some participants are sceptical about third-party certifiers, believing they might still try to satisfy or be pressured by the app developer (8\%), albeit in smaller numbers compared to first-party certifiers. 

Regarding community certifiers, some participants recognised the value of more diverse perspectives and knowledge, `out of the box' thinking and potential to offer unique solutions (5\%). 
Additionally, a number of participants valued the increased likelihood of spotting problems due to more reviewers (13\%).
In direct contrast, some participants, in particular of the \lowdetail\ variant, felt uncomfortable with having the code available to anyone to see, as this might include malicious actors (6\%).
In the \highdetail\ variant, we also see a rise in new type of concern that was not prevalent before, namely people questioning whether the open source community, in particular individuals, are qualified for the job, since they were not directly hired (5\%).

\subsection{Misconceptions and detail variants}\label{sec:user-study-misconceptions}
We identified two prominent misconceptions among participants in the \lowdetail\ variant. 
First, despite clear explanations in the introduction and comprehension questions, 18\% of participants incorrectly believed that certifiers could access or tamper with personal data. 
This misconception is particularly concerning when assessing comfort with community certifiers, as some mistakenly believed that their personal data would be visible to anyone. 
Second, as mentioned in \S\ref{sec:user-study-qualitative}, some users also believed that third-party certifiers are the only experts out of all (despite all certifiers being labelled as experts) or had superior access to the system (11\%).

To address these misconceptions in the \highdetail\ variant, we emphasised that all reviewers have an equal level of access to the system. 
We also changed the wording for L2 from reviewers having ``exclusive access to the code'' to them being ``directly authorised by the app developer'', as we noticed that the former wording led some participants to infer that third-party reviewers had higher access or expertise.
We also expanded the introductory explanation with details on concepts like Confidential Computing, its limitations, the importance of transparency, and the review process. 
To encourage deeper engagement with these explanations, we replaced the three True/False comprehension questions with ten multiple-choice questions.
As the \lowdetail\ variant already clarified that reviewers cannot access users' data, we made no changes to this aspect in the \highdetail\ variant.

These changes significantly reduced misconceptions in the \highdetail\ variant. 
The proportion of participants who believed reviewers had different access to user data dropped from 18\% to 6\%, while those suggesting differing expertise or access to the system completely disappeared, from previous 11\%. However, as mentioned before, some participants do believe it is potentially problematic that open source community was not explicitly hired or vetted for the job.
Overall, participants in the \highdetail\ variant demonstrated greater comfort with open-source certifiers, indicating that a clearer understanding of the system positively influences trust in higher levels of transparency.
\section{Related work}\label{sec:related-work}

The security of our HTTPS connections relies on domains sending their X.509 certificates to browsers, which then verify if the certificate is issued by an authorised Certificate Authority (CA). 
This mechanism inherently trusts CAs, creating a risk where compromised CAs could issue false certificates, leading to man-in-the-middle attacks \cite{diginotarmitm, verisignmitm, nicmitm}. 
To counter this, Google launched Certificate Transparency in 2013 \cite{laurie2013rfccertificate, laurie2014certificate}, which is currently mandatory in all major browsers \cite{ctenforcement, applesafaricertificatetransparency}.
Certificate Transparency allows public verification through three steps: CAs submit data to a transparency log, receive a signed inclusion proof, and incorporate it into the final certificate \cite{laurie2014certificate}.
Monitors (\S\ref{sec:custodians}) check these logs for anomalies and trigger revocation if needed.

In end-to-end encrypted communication, a similar issue arises with key directory servers.
CONIKS \cite{melara2015coniks}, inspired by Certificate Transparency, first introduced the concept of Key Transparency to ensure users receive the correct keys while protecting privacy.
Key Transparency has since evolved and found real-world applications including Google's open-source Key Transparency project \cite{googlekeytransparency}, along with implementations by WhatsApp and Apple \cite{whatsappkeytransparencyblog, whatsappkeytransparencywhitepaper, imessagekeytransparency}.

\label{sec:binary-transparency}
Binary Transparency \cite{binarytransparency}, addresses accountability in software supply chains and mitigates targeted attacks, ensuring the integrity of software from central repositories.
It is actively used in Google Pixel binaries \cite{androidbinarytransparency} and Go modules \cite{gobinarytransparency}.
Sigstore \cite{sigstore} is a collaborative project automating logging, signing, and monitoring for Binary Transparency. 
Similarly, Contour \cite{al2018contour}, MADT \cite{lins2023mobile}, and \textsc{CHAINIAC} \cite{nikitin2017chainiac} focus on software integrity through these methods.

Supply-chain Levels for Software Artifacts (SLSA)~\cite{linuxfoundation0} is a framework incorporating Binary Transparency concepts to counter software supply chain threats. 
It uses a tiered structure like ours, and can complement our approach to enhance the resilience of Confidential Computing systems. 
Our L2 and L3 aligns with SLSA v0.1 L4, emphasising reproducible builds and having a two-person review requirement through affiliated reporting first-party certifiers (Figure \ref{fig:reviewer-quadrant}, quadrant II).
Implementing our framework's L2 and L3 with hardened trusted builders (\S\ref{sec:variations}) achieves SLSA v1.0 L3.
Therefore, with specific implementation choices, our transparency framework can provide high levels of supply-chain security.
\section{Conclusions}\label{sec:conclusions}
This paper introduces the Confidential Computing Transparency framework designed to systematise transparency required by Trusted Execution Environments (TEEs) to reduce reliance on user trust (\S\ref{sec:transparency-levels}). 
The framework defines three progressive transparency levels, engaging first-party, third-party, and open-source reviewers.
It extends beyond existing industry practices by adding reviewer accountability and a robust trust chain with verifiable transparency logs, signed statements, and reproducible builds.
The tiered approach offers a practical approach for implementing trustworthy Confidential Computing systems across diverse scenarios.
We also address key questions about transparency in Confidential Computing, including its definition, importance, scope, beneficiaries, and facilitators (\S\ref{sec:defining-transparency}). 

To evaluate our framework's impact on end-user trust, we conducted a user study with over 800 participants.
Our user study has implications for academia and practitioners alike. 
Broadly, we find that participants favour higher transparency over unreviewed systems, and expressed varying comfort levels with different transparency levels depending on the type of personal data. 
Specifically, we find that third-party and open-source community reviewers are perceived as the most trustworthy, potentially making systems that incorporate these transparency levels more appealing to end-users. 
The primary reason that they are preferred over internal reviewing is that the latter is perceived as less reliable due to the perceived conflict of interest (\S\ref{sec:user-study-qualitative}). 

The study also finds that in the frequently occurring case in the industry where minimal information is provided, some misconceptions about transparency arise. 
For example, end-users may perceive open sourcing as less desirable fearing that it also means opening up the user data to everyone.
Similarly, end-users may believe that having the same exact component reviewed by third-party reviewers and community reviewers, the former has superior access to the component or expertise.
Our results suggest effectively communicating how Confidential Computing works, why transparency is important, and how reviews are made reduces such misconceptions.

Although we focused on Confidential Computing in this paper due to the unique \textit{attestability} of TEEs (\S\ref{sec:scope}), our transparency framework can also be applied to various Binary Transparency scenarios.
For binaries in remote TEEs, attestation ensures integrity, while in controlled environments users can directly validate binaries via signature checks and transparency log inclusion proofs. 
While we used software as our motivating example throughout, our approach is applicable to computer hardware too as long as it is within scope (\S\ref{sec:scope}).

\bibliographystyle{ACM-Reference-Format}
\bibliography{references}

\appendix

\section{Appendix}\label{apx:appendix}

\subsection{Ethical considerations}\label{apx:ethics}
Ethics approval was obtained from \textit{(anonymised for review)} prior to research. Given that our study involved human participants, several important ethical considerations were addressed:

\begin{itemize}
    \item \textbf{Informed consent:} We obtained informed consent from all users, with the option to withdraw from the study at any point or not answer any or all questions. We also provided direct contact with the researchers for any questions or concerns. 
    \item \textbf{Confidentiality and anonymity:} The data was anonymised and as such no single individual can be identified. We collected minimal demographics information, with all data analysed on an aggregate level.
    \item \textbf{Data Security:} All data was stored in electronic form on encrypted disks or a secure server. Only researchers involved in the project have direct access to raw data. Any publicly released data is anonymised and aggregated.
    \item \textbf{Compensation:} The participants were compensated for their time at the rate equivalent to the UK living wage per hour. The payment was made directly to the participant using the Prolific platform.
    \item \textbf{Risks to participants or researchers:} We did not foresee any risks or harms to either participants or the researchers involved in the study, or the greater public. 
\end{itemize}

\subsection{Open science}\label{apx:artefact}
To support open science and replicability of research, we make our user study artefacts available on (\textit{anonymised for review}). We include the full text of the user study, aggregated qualitative and quantitative data, demographic data and code for statistical analysis and plots. 

\subsection{Use of AI tools}
We used AI writing assistance tools, including OpenAI's ChatGPT and Anthropic's Claude, to refine this manuscript's language and clarity. 
These tools were used solely for stylistic editing and proofreading, with no contribution to the research content, data analysis, or scientific conclusions.

\subsection{User study script} \label{apx:user-study-script}

\minorsection{Welcome screen}
Thank you for participating in this study by \textit{(anonymised for review)}. Through this study we hope to learn more about people's preferences surrounding data sharing in different transparency scenarios. You will receive a payment of £3 for participating. The study is expected to take no more than 20 minutes.

Your participation is confidential and your identity will not be stored with your data. You will not be asked to provide any personally identifying information. Your answers will be used only for research purposes and in ways that will not reveal who you are. The results will be reported in aggregate form only, and cannot be identified individually. Any information that could identify you will be removed before data is shared with other researchers or results are made public. We will keep the raw data for the duration of the study and delete it by April 2025. For reproducibility, we may release a public appendix containing some aggregate and processed data.

You must be at least 18 years old to participate. By participating in this study, you consent to the data being used for this purpose. Your participation in this research is entirely voluntary and you have the right to withdraw consent at any time by closing your browser. For any questions during or after the study, please contact \textit{(anonymised for review)}.

\minorsection{Consent}
\noindent If you do not wish to participate in this study, please return your submission on Prolific by selecting the `Stop without Completing'. Are you happy to continue? \textit{(Note: Participant chooses between ``Continue with the study'' and ``Stop without completing''.)}

\minorsection{Background information}

Please watch the following informational video\footnotemark[\value{footnote}]. 
If you cannot watch it, click on the button below to see the text. 
You can proceed after the video finishes playing. 
\\

\textit{\minorsection{\textit{Low-detail} variant instructional script (written and video)}}
In this study, you will be presented with a number of imaginary apps. 
These apps run on a \textbf{Confidential Computing system} that aims to protect your data in such a way that even the app developer should not be able to see your data. 
However, this Confidential Computing system can only be trusted if it has been designed and built securely.

Imagine this system like a safe where your data is stored. 
The safe has a keypad, where you can enter your PIN to lock and unlock it. 
In order to trust the safe with protecting your data, you should trust that the keypad does not secretly record your PIN or have a special mechanism that opens the safe.

Sometimes, the Confidential Computing system can be checked by reviewers who certify that the product is safe to use. 
If their certification turns out to be faulty the reviewers can be held accountable.
The reviewers can be:
\begin{itemize}
    \item experts working for the app developer company
    \item experts who are granted exclusive access to the code for reviewing it. They may include consultants and auditors hired by the developer company, as well as regulatory authorities.
    \item experts from the broader software engineering community. The system is made publicly available for reviewing by academics, independent researchers, individuals who get rewards for finding bugs, or anyone who is interested in reviewing the code.
\end{itemize}
 
Or the system may not be reviewed at all.

Remember! Reviewers only look at the Confidential Computing system and do not see users’ personal data.
\\

\textit{\minorsection{\textit{High-detail} variant instructional script (written and video)}}
Many apps we use every day process our personal data in the cloud---a network of remote computers controlled by a cloud provider and the app developer. Apps protect data while it is stored or sent from one computer to another. However, data may become visible to the app developer and cloud provider while it's being processed, because some security protections must be disabled for processing.

Confidential Computing is a technology aimed at keeping our data private even while it is processed inside a computer. It does so by creating a special protected area within the computer, where data can be processed privately. This protected area keeps the data isolated from the rest of the computer, meaning that even the cloud provider or the app developer cannot see the data while it’s being processed.

However, Confidential Computing isn’t without limitations. Confidential Computing systems consist of multiple components, which need to work correctly together in order to provide privacy. If any part of the system has flaws or intentional weaknesses, it could allow the app developer, cloud provider, or another unauthorised person to see our private data.

One way to build confidence that a Confidential Computing system works correctly is to increase transparency. Transparency is an alternative to naively trusting that the system was built correctly and honestly. Transparency allows people to review how the system was built and publish their findings publicly. Though most people won’t review the system themselves, a transparent process allows expert reviewers to check it instead. By making the reviewers’ analysis and findings public, transparency can incentivise them to provide honest and good quality reviews.

These reviewers can be:
\begin{itemize}
    \item experts working for the app developer
    \item experts directly authorised by the app developer, such as hired consultants and auditors, as well as regulatory authorities
    \item experts from the broader software engineering community, such as academics, independent researchers, individuals who get rewards for finding bugs, or anyone who is interested in reviewing the code
\end{itemize}

Or, the Confidential Computing system may not be formally reviewed at all.

Remember! All of these reviewers get the same level of access to the system. This means that they all see how the reviewed components are constructed in order to validate whether they were built correctly. It’s important to note that reviewers cannot see, access or modify user data in any way.

\minorsection{Comprehension}

\textit{\minorsection{\textit{Low-detail} variant}}
\noindent Please answer the following questions. \textit{(Note: The answers are multiple choice between True / False)}
\vspace{0.5em}

\begin{enumerate}[label=\textbf{Q\arabic*}, itemsep=0pt, parsep=0pt, topsep=0pt, partopsep=0pt]
    \item A Confidential Computing system claiming to be secure is always secure, \textbf{regardless of how it was built}.
    \item  Reviewers who certify the system's safety \textbf{can be held accountable} if their certification is faulty.
    \item Reviewers get access to the full system, \textbf{including users' personal data}.
\end{enumerate} 

\newpage

\textit{\minorsection{\textit{High-detail} variant}}
\noindent Please answer the following questions. \textit{(Note: The answers are multiple choice)}
\vspace{0.5em}

\begin{enumerate}[label=\textbf{Q\arabic*}, itemsep=0pt, parsep=0pt, topsep=0pt, partopsep=0pt]
    \item Why is our data more at risk while it’s being processed, especially in the cloud?
    \begin{enumerate}
        \item Because data protection is temporarily disabled while it’s being processed
        \item Because cloud providers do not provide any security measures
        \item Because the data is typically not protected when it is stored or sent from one computer to another
        \item Because cloud providers and app developers always have full access to user data
    \end{enumerate}
    \item  What is the main purpose of Confidential Computing, and how does it protect data?
    \begin{enumerate}
        \item To store data permanently in a secure location on a computer
        \item To protect data when it’s being sent from one computer to another
        \item To keep data private during processing by creating a protected area where it can be processed privately
        \item To enhance computer processing speed by isolating certain data
    \end{enumerate}
    \item How can Confidential Computing fail to protect your data?
    \begin{enumerate}
        \item Confidential Computing never fails to protect data
        \item If a part of the Confidential Computing system fails to work correctly
        \item If there is not enough data
        \item If the user forgets their password
    \end{enumerate}
    \item How can transparency help build confidence in Confidential Computing? (Select all that apply)
    \begin{enumerate}
        \item By making the system available for review, it allows experts to check that the system was built correctly
        \item Transparency means that all users must review the system themselves to ensure it is secure
        \item Transparency can incentivise expert reviewers to provide good quality and honest reviews by making their findings public
        \item Transparency reduces the need for any review, as users can trust that the system is secure by design
    \end{enumerate}
    \item[\textbf{Q5 - 8}]  Match the GIFs to reviewer types. \textbf{\texttt{[GIFS AND DESCRIPTIONS OF THE FOUR TRANSPARENCY LEVELS]}}
    \setcounter{enumi}{8}
    \item Which of the following reviewer categories gets a higher level of access to the system compared to others?
    \begin{enumerate}
        \item Experts from the broader software engineering community, such as academics, independent researchers, individuals who get rewards for finding bugs, or anyone who is interested in reviewing the code
        \item Experts directly authorised by the app developer, such as hired consultants and auditors, as well as regulatory authorities
        \item Experts working for the app developer
        \item All of them get the same level of access

    \end{enumerate}
    \item Do reviewers gain access to user data?
    \begin{enumerate}
        \item Yes
        \item No
    \end{enumerate}
    
\end{enumerate} 

\minorsection{Instructions}

You will now be presented with a number of \textbf{different multi-purpose virtual assistant apps} that help you with your everyday tasks. 
To operate efficiently, each app requires access to \textbf{different types of your data}. 
Each app runs on the previously described \textbf{Confidential Computing system}. 
The apps have been developed by a startup and are just about to be released.

For each of the scenarios, you will be asked about your comfort and confidence using the app under the specified conditions.

\minorsection{Core study}
(\textit{Note: Six versions of the following screen are presented to the participants, varying the data types and transparency level descriptions as listed after the questions.})

Virtual assistant app \textbf{\texttt{[LETTER IN THE RANGE A-F]}} requires permission from you to access your \textbf{\texttt{[DATA TYPE, TABLE \ref{tab:data-types}]}}. This includes \textbf{\texttt{[EXAMPLES OF DATA TYPE, TABLE \ref{tab:data-types}}]}.

In order to assess if the Confidential Computing system is designed and built securely, the system was \textbf{\texttt{[TRANSPARENCY LEVEL, TABLE \ref{tab:level-descriptions}]}}.

\vspace{0.5em}

\noindent(\textit{Note: Participants answer each question using a slider ranging from 0 to 100.})

\begin{enumerate}[label=\textbf{Q\arabic*}, itemsep=0pt, parsep=0pt, topsep=0pt, partopsep=0pt]
    \item How comfortable would you feel using this app?
    \item How confident would you feel about this app’s ability to keep your \textbf{\texttt{[DATA TYPE]}} data safe?
\end{enumerate}

\begin{table}[ht!]
\centering
\footnotesize
\begin{tabular}{@{}lp{5.1cm}@{}}
\textbf{Data type} & \textbf{Selected examples} \\
\hline
Lifestyle-behavior & Religious and political beliefs, interests and preferences, browsing habits, family and relationships. \\
Social-economic & Ethnicity, physical characteristics (from pictures and videos), age and gender, professional career. \\
Tracking & Email address, physical address, phone number, phone recordings and text messages, geolocation, IP address. \\
Financial & Credit history, financial assets, bank information and credit card numbers, purchases, transactions, taxes. \\
Authenticating & Passwords, passport data, government ID data, usernames. \\
Medical-health & Personal health history and diagnoses, prescriptions, mental health records, disabilities, genetic data. \\
\end{tabular}
\caption{Six data type categories and their selected examples from Chua et al. \cite{chua2021effects}}
\label{tab:data-types}
\end{table}

\begin{table}[!ht]
\centering
\footnotesize
\begin{tabular}{p{0.11\textwidth} p{0.315\textwidth}}
\textbf{Transparency Level} & \textbf{Text in the script} \\ 
            & \textit{In order to assess if the Confidential Computing system is designed and built securely, the system was reviewed by ... }\vspace{1mm}\\
\textbf{L1} &  ... experts working for the app developer company. \\ \hline
\textbf{L2} & ... experts who are granted exclusive access to the code for reviewing it. They may include consultants and auditors hired by the developer company, as well as regulatory authorities. \\ \hline
\textbf{L3} & ... experts from the broader software engineering community. The system is made publicly available for review by academics, independent researchers, individuals who get rewards for finding bugs, or anyone interested in reviewing the code. \\ \hhline{==}
\textbf{No transparency} & The Confidential Computing system was not formally reviewed. \\
\end{tabular}
\caption{Transparency levels and their corresponding descriptions in the user study script for \lowdetail\ variant.}
\label{tab:level-descriptions}
\end{table}

\begin{table}[!ht]
\centering
\footnotesize
\begin{tabular}{p{0.11\textwidth} p{0.315\textwidth}}
\textbf{Transparency Level} & \textbf{Text in the script} \\ 
            & \textit{In order to assess if the Confidential Computing system is designed and built securely, the system was reviewed by ... }\vspace{1mm}\\
\textbf{L1} &  ... experts working for the app developer. \\ \hline
\textbf{L2} & ... experts directly authorised by the app developer, such as hired consultants and auditors, as well as regulatory authorities. \\ \hline
\textbf{L3} & ... experts from the broader software engineering community, such as academics, independent researchers, individuals who get rewards for finding bugs, or anyone who is interested in reviewing the code. \\ \hhline{==}
\textbf{No transparency} & The Confidential Computing system was not formally reviewed. \\
\end{tabular}
\caption{Transparency levels and their corresponding descriptions in the user study script for \highdetail\ variant.}
\label{tab:level-descriptions}
\end{table}

\minorsection{Additional questions}
\textit{Note: Participants answer each question using a slider ranging from 0 to 100.}

\begin{enumerate}[label=\textbf{Q\arabic*}, itemsep=0pt, parsep=0pt, topsep=0pt, partopsep=0pt]
    \item How would you best describe your approach to purchasing or experimenting with new technology?
    \item How likely are you to use a multi-purpose virtual assistant app (e.g. Siri, Alexa, Google Assistant)?
    \item How concerned are you about other people getting access to your data without your consent?    
    \item Imagine your data was leaked or tampered with. How worried would you be for each of the data types listed below? 
    \item Regardless of the data type, how comfortable would you feel with the virtual assistant app being reviewed with each of the following options?
\end{enumerate}

\vspace{0.5em}
\noindent \textit{Note: The following is an open-ended question with a text box for the response.}

\begin{enumerate}[label=\textbf{Q\arabic*}, start=11, itemsep=0pt, parsep=0pt, topsep=0pt, partopsep=0pt]
    \item Please explain how you assigned the scores to the review options. 
\end{enumerate}

\minorsection{Demographics}
\textit{Note: We collect the following information: age, gender, level of schooling, field of work/study and country in which the participants lived the longes. 
We omit the full demographics questions in the interest of space.}

\minorsection{End of survey and payment}
Thank you for taking part in this study. 
Please click on the `Finish' button below to be redirected back to Prolific and register your submission.
(Note: \textit{Upon clicking the `Finish' button, the participants are redirected to Prolific platform and payment is made.})

\newpage

\subsection{User study demographics} \label{apx:demographics}

\begin{table}[ht!]
\centering
\footnotesize
\begin{tabular}{@{}ll@{}}
\toprule
\textbf{Characteristic} & \textbf{Distribution (\%)} \\ \midrule
\textbf{Age}            &                             \\
\hspace{10pt} 18-24     & 21\%                      \\
\hspace{10pt} 25-34     & 39\%                      \\
\hspace{10pt} 35-49     & 27\%                      \\
\hspace{10pt} 50+       & 13\%                       \\\midrule
\textbf{Gender}         &                             \\
\hspace{10pt} Female    & 44\%                      \\
\hspace{10pt} Male      & 54\%                      \\
\hspace{10pt} Other     & 2\%        \\ \midrule
\textbf{Education}      &                             \\
\hspace{10pt} High School & 22\%                    \\
\hspace{10pt} Trade School & 12\%                   \\
\hspace{10pt} University (Bachelor's) & 45\%       \\
\hspace{10pt} University (Master's/PhD) & 21\%     \\  \midrule
\textbf{Field of Study} &                             \\
\hspace{10pt} STEM       & 29\%                    \\
\hspace{10pt} Economics/Finance/Business        & 18\%   \\
\hspace{10pt} Health/Medicine & 10\%               \\
\hspace{10pt} Arts/Humanities & 10\%               \\
\hspace{10pt} Social Sciences & 8\%               \\
\hspace{10pt} Trades/Personal services    & 5\%       \\
\hspace{10pt} Other/Prefer not to say   & 20\%  \\ \midrule
\textbf{Region}         &                             \\
\hspace{10pt} Europe     & 34\%                    \\
\hspace{10pt} North America   & 33\%                    \\
\hspace{10pt} Middle East and Africa & 17\%       \\
\hspace{10pt} South America   & 8\%                    \\
\hspace{10pt} Asia and Pacific & 8\%               \\ \bottomrule
\end{tabular}
\caption{Demographics of user study participants.}
\label{tab:demographics}
\end{table}

\end{document}